\documentclass{article}

\usepackage{PRIMEarxiv}

\usepackage[utf8]{inputenc} % allow utf-8 input
\usepackage[T1]{fontenc}    % use 8-bit T1 fonts
\usepackage{hyperref}       % hyperlinks
\usepackage{url}            % simple URL typesetting
\usepackage{booktabs}       % professional-quality tables
\usepackage{amsfonts}       % blackboard math symbols
\usepackage{nicefrac}       % compact symbols for 1/2, etc.
\usepackage{microtype}      % microtypography
\DisableLigatures{encoding = *, family = *}
\usepackage{amsmath}        % for \lVert and \rVert
\usepackage{lipsum}
\usepackage{fancyhdr}       % header
\usepackage{graphicx}
\usepackage{array}
\usepackage{booktabs}
\usepackage{multirow}
\usepackage{tabularx}
\usepackage{adjustbox}
\usepackage{setspace}
\graphicspath{{media/}}     % organize your images and other figures under media/ folder

\title{Advancing credit mobility through stakeholder-informed AI design and adoption
}

\author{
  Yerin Kwak,
  Siddharth Adelkar,
  Zachary A. Pardos$^{*}$ \\
  Berkeley School of Education, University of California, Berkeley, CA \\
  \texttt{\{kwak, adelkar, pardos\}@berkeley.edu} \\
  $^{*}$Corresponding author
}

\begin{document}
\maketitle

\begin{abstract} %1920 characters
Transferring from a 2-year to a 4-year college is crucial for socioeconomic mobility, yet students often face challenges ensuring their credits are fully recognized, leading to delays in their academic progress and unexpected costs. Determining whether courses at different institutions are equivalent (i.e., articulation) is essential for successful credit transfer, as it minimizes unused credits and increases the likelihood of bachelor’s degree completion. However, establishing articulation agreements remains time- and resource-intensive, as all candidate articulations are reviewed manually. Although recent efforts have explored the use of artificial intelligence to support this work, its use in articulation practice remains limited. Given these challenges and the need for scalable support, this study applies artificial intelligence to suggest articulations between institutions in collaboration with the State University of New York system, one of the largest systems of higher education in the US. To develop our methodology, we first surveyed articulation staff and faculty to assess adoption rates of baseline algorithmic recommendations and gather feedback on perceptions and concerns about these recommendations. Building on these insights, we developed a supervised alignment method that addresses superficial matching and institutional biases in catalog descriptions, achieving a 5.5-fold improvement in accuracy over previous methods. Based on articulation predictions of this method and a 61\% average surveyed adoption rate among faculty and staff, these findings project a 12-fold increase in valid credit mobility opportunities that would otherwise remain unrealized. This study suggests that stakeholder-informed design of AI in higher education administration can expand student credit mobility and help reshape current institutional decision-making in course articulation.
\end{abstract}

% keywords can be removed

\section{Introduction}
Transfer pathways from two-year institutions are a major contributor to enrollments at four-year institutions in the US \cite{velasco2024tracking}. Previous research has highlighted significant inequalities in access to higher education \cite{stoet2020gender,jackson2020century}, underscoring the importance of these pathways. A central barrier—yet an essential component—is the course equivalency evaluation, which determines whether a course at one institution is sufficiently similar to another course at another institution for credits to be transferred. Establishing course equivalency between institutions is essential to enabling successful credit transfer~\cite{belfield2017really,bender1990spotlight,donovan1987transfer,knoell1990transfer,boggs2010growing} and increasing the likelihood of bachelor’s degree completion~\cite{hodara2016improving,monaghan2015community,hodara2017exploring}. Yet, students lose 43\% of their college credits on average when transferring~\cite{us2017higher}, leading to delays in their academic progress and unexpected costs~\cite{belfield2017really,logue2024possible}.
% https://www.insidehighered.com/news/quick-takes/2025/02/24/survey-finds-widespread-challenges-credit-transfer (the higher education consulting firm Sova and the Beyond Transfer Policy Advisory Board, 2025, More than half of respondents who attempted a transfer reported losing at least some credits in the process, Meanwhile, a fifth of respondents who tried to transfer credits had to repeat classes they’d already taken)

% https://upcea.edu/survey-3-in-10-students-lose-significant-academic-credits-transferring-between-colleges/ (UPCEA, 30% of students lose at least a quarter of existing academic credit , 2022)

Various states in the US have adopted different statewide course equivalency policies to minimize unused credit accumulation and improve the efficiency of course agreement processes~\cite{ecs2022transfer,ignash2000evaluating}. One such approach is course articulation, the process of developing formal agreements to ensure that courses from a “sending” institution are accepted as comparable and transferable at a “receiving” institution. As of November 2025, there are 23 states, including California, Maryland, and Ohio, that have implemented a statewide course articulation system. 

% do not start with however -- little bit abrupt (little bit of movitation/background) 
In the current course articulation process, a request for transfer recognition requires faculty and articulation staff to review all candidate courses and determine whether to accept, modify, or deny the request \cite{ciac2013handbook}. This process demands substantial time from faculty and staff, as well as significant financial resources. Re-articulating courses between the 115 degree-granting California Community Colleges and the nine University of California (UC) campuses would require evaluating at least 63 million equivalencies \cite{pardos2019data}. The task becomes intractable when expanding to other postsecondary segments, private or out of state schools, and accommodating course and new program changes (e.g., the emergence of Data Science). Around 70\% of articulation experts surveyed identify this excessive workload as a major challenge in their transfer credit evaluation work~\cite{xu2023convincing}. Additionally, studies have highlighted another challenge in this manual review process, noting that it can sometimes lead to inconsistencies due to the reliance on individual experiences and interpretations of credit acceptance norms~\cite{barry1992establishing}. The financial investments in strictly manual processes are also considerable. For example, Alabama invested \$1,200,000 in the startup of their transfer and articulation program in 1994, followed by additional funding to maintain and improve the system as new courses and institutions were introduced. From 1998 to 2008, the state spent \$500,000 annually to support these ongoing improvements~\cite{katsinas2016alabama}.

%de2023psychological- targeted interventions that address these concerns can alleviate resistance to AI tools. 
To address these burdens, recent efforts have explored the use of artificial intelligence (AI) to support this work~\cite{pardos2019data,xu2023convincing}. However, a central question remains how such systems should be designed to address concerns about AI, thereby fostering trust and effective use among faculty and staff~\cite{de2023psychological}. Prior studies have documented that human decision-makers tend to discount algorithmic recommendations more heavily than equivalent recommendations made by humans~\cite{dietvorst2015algorithm}. This is especially pronounced among experts and in socially consequential contexts such as higher education, governance, and legal decision-making~\cite{filiz2021tragedy}. Explanations for this include a tendency to lose trust when algorithms make visible errors~\cite{highhouse2008stubborn} and a belief that algorithms cannot learn and improve in the way humans do~\cite{dietvorst2015algorithm}. However, studies also indicate that AI systems designed to reflect human preferences~\cite{russell2019human} and achieve \textit{cognitive compatibility} can facilitate augmented decision making and trust building, potentially leading to increased acceptance of algorithmic recommendations~\cite{burton2020systematic,bohlen2025overcoming}. Thus, identifying stakeholders’ expectations and concerns and translating them into concrete design requirements is critical for improving the trustworthiness and eventual adoption of AI in course articulation.

\begin{figure*}[!htbp]
\centering
\includegraphics[width=1\textwidth]{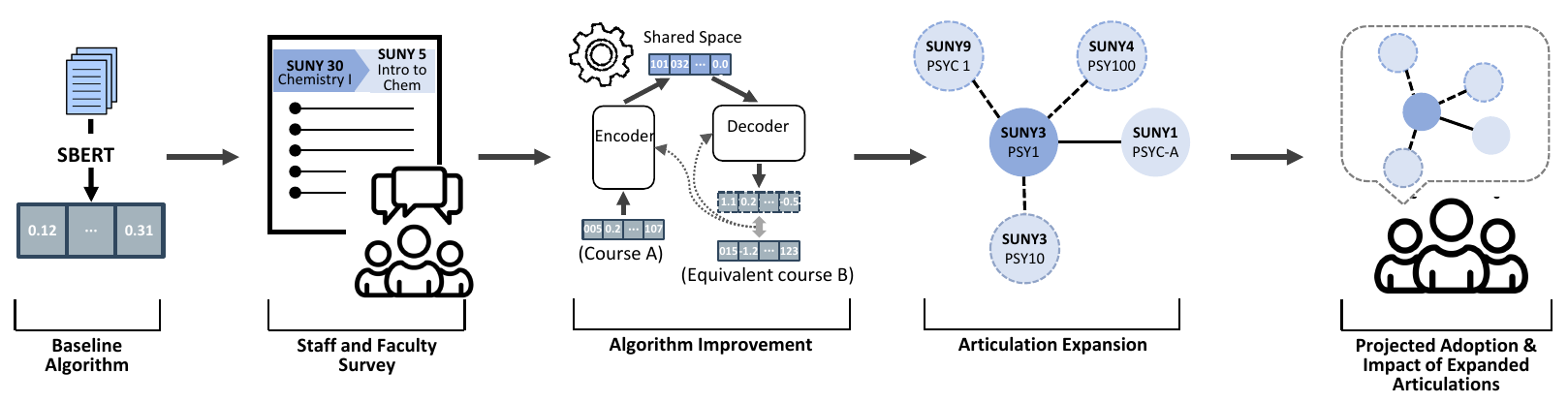}
\caption{\textbf{Overview of the study workflow.} This figure illustrate the end-to-end workflow of the study, beginning with generating articulation recommendations using SBERT embeddings, followed by faculty and staff evaluations of the recommendations. Feedback from the evaluation informs advancements to the method. The improved approach is used for articulation expansion, and we project adoption rates for these expanded articulation candidates based on observed faculty and staff acceptance patterns.}
\label{fig:fig0}
\end{figure*}

This study (outlined in Figure \ref{fig:fig0} ) applies AI to define and maintain real-world course articulation scenarios in one of the largest postsecondary systems in the US, the State University of New York system (SUNY). We first examine how faculty and staff across New York State institutions perceive AI-assisted articulation recommendations generated by a baseline model. Building on previous research that used student enrollment histories and course descriptions for algorithmic course articulation~\cite{pardos2019data}, and iterating our approach based on stakeholder feedback, we introduce a substantially more accurate model, owing in part to the power of the current generation of Natural Language Processing (NLP). Using this enhanced model, we identify previously undiscovered articulation pairs among 120,749 transferable courses offered across SUNY and estimate the expected articulation agreement expansion based on faculty and staff adoption patterns observed with the baseline recommendations. This study contributes to the emerging areas AI in higher education and academic pathways research \cite{kizilcec2023pipelines}.

%This study applies AI to the course articulation process and presents a methodology that reflects the perspectives and design requirements of faculty and staff, achieving substantially higher accuracy than prior approaches and revealing numerous new credit mobility pathways that would otherwise have remained unrealized.

\section{Background}
%SUNY
The SUNY system is the largest comprehensive university system in the US, serving nearly 1.3 million students across 64 institutions, including 14 University Centers, 13 University Colleges, 7 Technology Colleges, and 30 Community Colleges. The first three campus types are bachelor-granting (four-year) institutions, whereas the community colleges offer two-year programs. Transfer plays a central role in the system: More than 45\% of SUNY bachelor's degree recipients begin as transfer students. Currently, 32.08\% of SUNY students transfer to another institution, with 52.08\% of these students transferring to a SUNY institution. In this context, SUNY aims to improve its transfer and articulation processes by reducing credits loss. Its credit acceptance rate is currently 64-67\%, above the national average of 56-60\%, with a goal of reaching 80\% \footnote{\url{https://transfer.suny.edu/task-force/}}.

\subsection{Datasets}
\begin{figure*}[!htbp]
\centering
\includegraphics[width=0.7\textwidth]{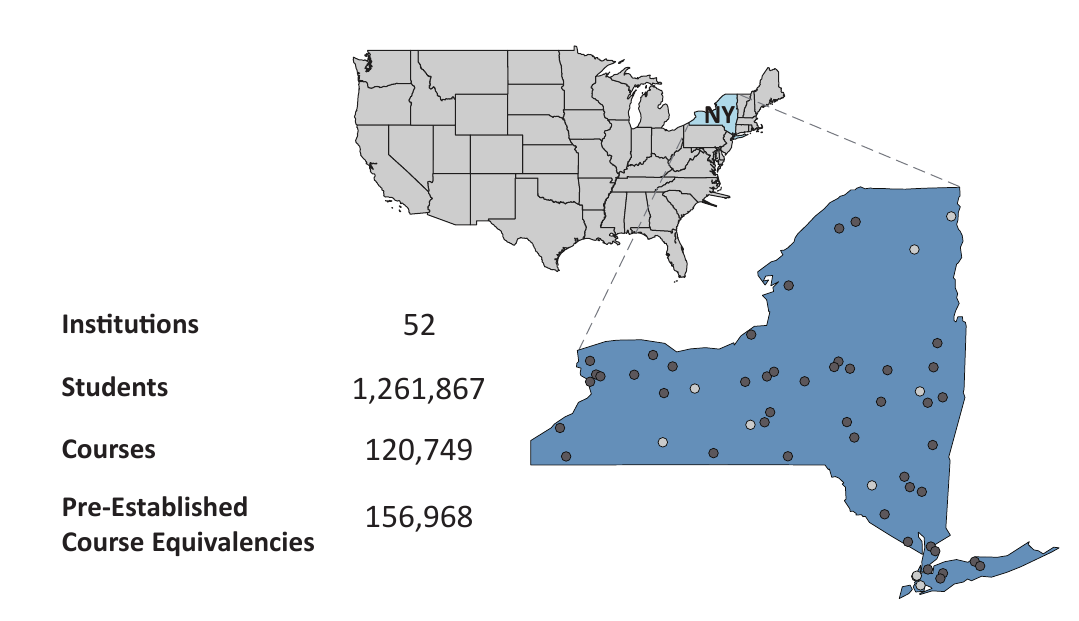}
\caption{\textbf{Overview of data sources.} Geographic distribution of institutions and dataset descriptives. Dark circles represent colleges included in this study, while white circles indicate those not included.}
\label{fig:fig2}
\end{figure*}

A total of 52 SUNY institutions provided complete data, including course long titles and descriptions. The dataset consists of 1,261,867 students, 120,749 transferable courses with their titles and descriptions, and 156,968 articulation pairs formally established by SUNY. Because the system includes both four-year and two-year campuses, these articulation pairs span multiple transfer pathways. Among them are two-year to two-year articulations (18,769 pairs), which often reflect students pursuing alternate academic options within the community college sector; four-year to two-year articulations (25,571 pairs), including reverse transfer, in which students transfer credits earned at a SUNY four-year institution back to a SUNY community college to complete an associate degree; four-year to four-year articulations (53,667 pairs), a lateral route typically used by students seeking new academic opportunities or institutional environments; and two-year to four-year articulations (58,961 pairs), the most common vertical transfer pathway enabling students to apply community college credits toward a bachelor’s degree. An overview of the data sources is provided in Figure \ref{fig:fig2}.

% FIG 2

\section{Staff and Faculty Survey}
\label{sec:survey}
Manually determining and maintaining comprehensive course articulations among the thousands of institutions in the U.S. is intractable at scale. Prior research has demonstrated the potential of computational approaches to streamline this process, introducing a method for identifying course equivalencies between 2-year community colleges and 4-year universities in California \cite{pardos2019data}. This approach embeds courses from an institution into a semantic vector space based on student enrollment sequences, similar to how words are embedded based on their contexts in a corpus \cite{mikolov2013efficient}. The courses can then be translated to other institutions' vector spaces using knowledge of a limited set of existing articulations and then compared to one another to estimate articulatability. In practice, however, articulation decisions are still made manually by faculty and articulation staff. The limited adoption of such AI systems suggests that successful implementation depends not only on their objective performance but also on how they are subjectively perceived by key stakeholders \cite{de2023psychological}. Understanding these perceptions and designing targeted interventions to address their concerns are therefore important for supporting the adoption of AI-assisted tools.

\subsection{Survey Design}

We began by examining faculty and staff perceptions of the recommendations generated using a baseline algorithm (SBERT;~\cite{reimers2019sentence}) and their adoption rates, and by identifying areas for algorithmic development based on their feedback. The typical workflow for establishing articulation agreements begins with the sending institution identifying courses for which it seeks transfer recognition and providing documentation outlining course content. Staff and faculty at the receiving institution then review this material and compare it against their own course offerings. If a course is deemed equivalent, an articulation agreement is approved \cite{ciac2013handbook, xu2023convincing}. Based on this workflow, the surveys presented staff and faculty with a hypothetical scenario in which an AI recommends a set of potential course equivalencies and asked whether they would accept any of these recommendations. For the algorithm, we selected SBERT because it was an advanced sentence-embedding model available at the time the first survey was conducted. Section~\ref{result-prediction} reports the performance of this model in predicting articulations. Using course titles and descriptions, we generated course vectors with SBERT, and identified recommended courses as those with the highest cosine similarity to the sending institution’s course vector.  Two surveys were conducted: the first was administered to articulation staff from the New York State Transfer and Articulation Association, and the second to faculty members within SUNY.

In the first study, 41 articulation staff members from the New York State Transfer and Articulation Association participated \cite{xu2023convincing}. We constructed the evaluation scenarios using courses from Laney Community College (sending institution) and UC Berkeley (receiving institution). Ten sending-institution courses were randomly sampled with replacement to form each of five sets, resulting in 45 unique courses across all sets. Each participant was assigned one of the five sets using a round-robin procedure to ensure that all sets were evaluated a comparable number of times. For each of the ten scenarios in a set, participants were shown seven recommended courses and asked to select the one they considered an appropriate articulation match, or indicate that no suitable match was present. Seven options were chosen to balance providing sufficient choice without overwhelming participants. We calculated the adoption rate as the proportion of scenarios in which participants selected one of the recommended courses as an appropriate match. Participants were also invited to provide open-ended feedback on the recommendations.

Following the staff survey, we also conducted a study with faculty, who typically make the final articulation decisions and may hold different perspectives. Faculty are often asked to make determinations about the credit equivalency of their own courses or other courses in their department or proximal subject area. Therefore, unlike the staff survey, we showed course articulation recommendation results from the faculty participant's own institution, allowing for a more ecological valid evaluation. The faculty survey otherwise followed the same design and procedure as the staff survey. As in the staff study, participants were given seven recommended courses per scenario and asked to select the one they considered an appropriate match or indicate that no suitable match was present. A total of 19 SUNY faculty participated, each evaluating eight scenarios.

\subsection{Adoption Patterns and Feedback}
Table \ref{tab:survey} shows the results from the two studies, including adoption rates and qualitative feedback. In the staff survey, the overall articulation recommendation adoption rate was 63.25\%. Qualitative feedback highlighted concerns about low accuracy, particularly that simple word or phrase matching was insufficient for establishing true articulations. For example, one participant noted that a course titled \textit{Topics in Algebra} was recommended as equivalent to \textit{Elements of Algebra}, likely because both included the word “\textit{algebra},” even though the courses could cover different content. Similarly, the algorithm recommended \textit{South Asian American Historical and Contemporary Issues} as equivalent to \textit{Asian-American Communities} seemingly due to overlapping terms in the descriptions. For instance, both mention “\textit{Asian American communities}” and “\textit{social, political and economic contexts.}” However, staff noted that the former course has a narrower regional focus and therefore cannot be considered equivalent. These comments underscore the need for algorithms that go beyond surface-level lexical similarity to capture deeper course content.

In the faculty survey, the recommendation adoption rate was 59.21\%. Faculty raised concerns similar to those expressed by staff, pointing out that the system appeared to rely too heavily on text matching when generating recommendations. They observed that courses were often suggested as equivalent simply because their titles or descriptions contained overlapping words, even when the actual content could differ substantially. They also emphasized that academic language is jargon-heavy, and that identical terms may not carry the same meaning across institutions. 

%table here
\begin{table}[htbp]
\centering
\caption{\textbf{Staff and faculty survey findings.} This table summarizes acceptance rates and example qualitative feedback.}
\label{tab:survey}

\renewcommand{\arraystretch}{1.3}
\setlength{\tabcolsep}{6pt}

\begin{tabularx}{\textwidth}{
    p{3cm}
    >{\centering\arraybackslash}X
    >{\centering\arraybackslash}X
}
\toprule
 & \textbf{Staff Survey} & \textbf{Faculty Survey} \\
\midrule
\textbf{\# Courses / \# Scenarios} &
7 courses / 10 scenarios &
7 courses / 8 scenarios \\
\midrule
\textbf{Model} &
SBERT (no enrollment) &
SBERT (no enrollment) \\
\midrule
\textbf{\# Participants} &
41 articulation staff (New York State Transfer and Articulation Association) &
19 SUNY faculty \\
\midrule
\textbf{Acceptance Rate} &
63.25\% &
59.21\% \\
\midrule
\textbf{Comments} &
\parbox[t]{\hsize}{\raggedright\small
``A course titled \textit{Topics in Algebra} at the sending institution and one titled \textit{Elements of Algebra} at the receiving institution may look similar, but they could be entirely different.''

\vspace{0.9em}

``The sending institution’s course\textsuperscript{†} covers Asian-American communities more broadly, while the receiving institution’s course\textsuperscript{‡} focuses specifically on South Asian American communities.''
} &
\parbox[t]{\hsize}{\raggedright\small
``The system seems to match courses based on titles and similar words in the course descriptions. The platform needs more training to understand that courses are not actually equivalent even if their titles and descriptions share similar words. The system needs to be checked by a human for accuracy.''

\vspace{0.9em}

``The AI would have to be able to do more than just look for common vocabulary between course descriptions. Academic language is often dense, opaque, and jargon-heavy. That jargon is used in specific ways on specific campuses, and may not translate equivalently to another campus’s jargon (e.g., Concurrent enrollment at NCC is what some other campuses call Dual Enrollment, which is a different program). Because of this variance and opacity, I would not trust the AI’s percentage of confidence in its suggestions, nor would I trust someone with less experience to be able to make a sound decision based on that percentage score.''
} \\
\bottomrule
\end{tabularx}

\vspace{0.4em}
\begin{flushleft}
{\fontsize{6.2}{6.5}\selectfont
\mbox{\textsuperscript{†}\,Asian-American Communities: Study of political, economic, and social structures of Asian-American communities, past and present, with emphasis on current issues and problems.} 

\mbox{\textsuperscript{‡}\,South Asian American Historical and Contemporary Issues: Development of South Asian American communities within the social, political and economic contexts of South Asia and the U.S.}
}
\end{flushleft}

\end{table}

\subsection{Design Implications for Algorithmic Improvement}
The average adoption rate was 61.23\%, with staff and faculty showing comparable rates. This indicates a promising starting point, though nearly four out of ten scenarios still involved participants rejecting all recommended courses. This underscores the need for more accurate outputs. Open-ended feedback offered guidance on the development of a methodology for algorithmic improvement. The most common issues identified was the need for more precise recommendations: algorithms should generate results that accurately capture genuine content alignment rather than relying on simple word or phrase matching, and can discern nuanced semantic and contextual differences, even when courses are offered at different institutions and contain institution-specific language. 

To refine our algorithm in response to these limitations, we adopted the Multilingual Pseudo-Supervised Refinement (MPSR) approach, which is used to refine a shared embedding space to improve the alignment of multilingual embeddings~\cite{chen2018unsupervised}. Prior work has shown that, when applied after an initial alignment, MPSR not only predicts word translations more accurately but also produces embedding spaces in which cross-lingual word similarities correlate more strongly with human semantic judgments than in earlier bilingual alignment models. These findings suggest that MPSR enhances the semantic alignment of embeddings across languages, possibly by drawing on information from multiple languages~\cite{chen2018unsupervised}. This makes it a suitable choice for our setting, where institutions play an analogous role to languages: by mapping courses titles and descriptions from different institutions into a shared space, MPSR can strengthen semantic similarity rather than shallow textual similarity.

\section{Algorithm Improvement}
\subsection{Methodological Framework}
\begin{figure*}[!htbp]
\centering
\includegraphics[width=0.95\textwidth]{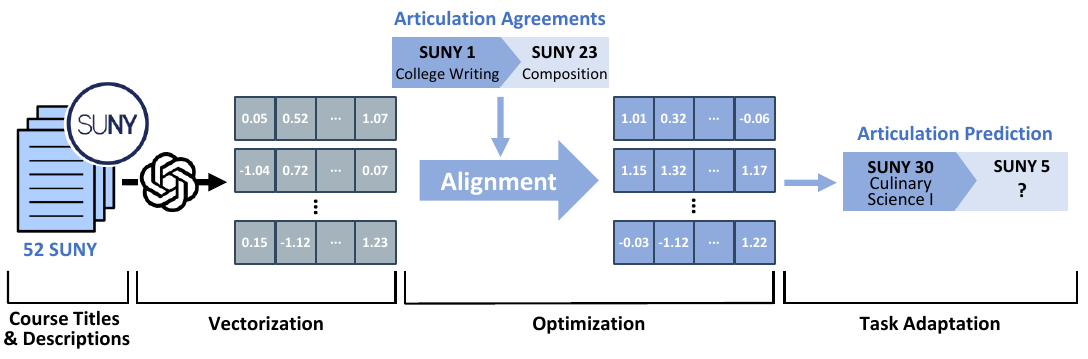}
\caption{\textbf{Methodological framework for AI-assistive course equivalency process.} This figure provides a high-level overview of the methodology and the task objectives. The process begins with the vectorization of course titles and descriptions using a readily available state-of-the-art large language model. We then refined these vectors through a shared space alignment technique, leveraging pre-established course articulations. The resulting vectors were validated using faculty-established articulations and then applied to system-specific tasks (i.e., course articulation at SUNY), with the outcomes were evaluated accordingly.}
\label{fig:fig1}
\end{figure*}

We applied our AI course equivalency framework (Fig. \ref{fig:fig1}) to a major real-world task in higher education: course articulation within the State University of New York (SUNY) system. The methodological framework uses course titles and catalog descriptions, which are vectorized and then aligned using our faculty- and staff-informed technique. We then apply a K-nearest neighbor approach (KNN;~\cite{fix1985discriminatory}) to predict candidate articulations and evaluate performance using recall@1 and recall@5, based on existing articulations established by SUNY faculty. This case illustrates course articulation challenges in U.S. higher education systems. SUNY represents the largest comprehensive system with 64 member institutions, including universities and community colleges \cite{suny2020history}.

% FIG 1

\subsection{Shared Space Alignment (SSA)}
Building on insights from the surveys, we developed a supervised alignment technique, Shared Space Alignment (SSA). Our methodological objective was to develop vector representations of courses that position similar courses close together in a vector space, allowing the identification of course equivalencies to become a nearest neighbor search problem. To encourage similar courses from different campuses to cluster more closely in spite of superficial differences in their descriptions, we propose Shared Space Alignment (SSA), a technique inspired by a machine translation algorithm called MPSR that is used to align vector spaces of multiple languages \cite{chen2018unsupervised}. However, unlike MPSR, which optimizes embeddings toward a shared space, SSA instead shifts the optimization toward the destination space. This better reflects the real-world practice of course matching based on the receiving institution side and results in empirical gains in recall and explainability. 

\begin{figure*}[!htbp]
\centering
\includegraphics[width=0.7\textwidth]{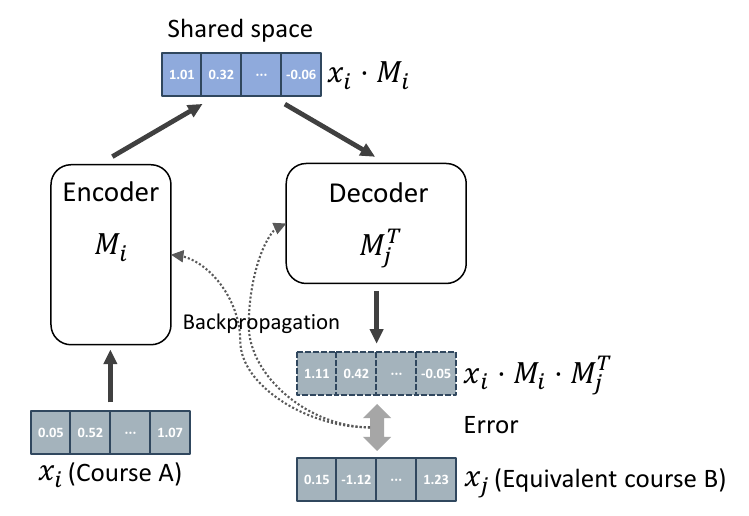}
\caption{\textbf{The encoder and decoder process of SSA.} A course vector $x_i$ from college $i$ is encoded into a shared space using the transformation matrix $M_i$.The resulting vector, $x_i \cdot M_i$, is then decoded into the vector space of college $j$ using the transpose of the transformation matrix $M_j^{T}$. The error is calculated between this predicted vector, $x_i \cdot M_i \cdot M_j^{T}$, and the true equivalent course vector $x_j$ from college $j$. This error is minimized through iterative updates of the transformation matrices $M$ using backpropagation \cite{rumelhart1986learning}. After optimization, the outputs of this algorithm are blue encoded shared space vectors, $x_i \cdot M_i$.}
\label{fig:fig4}
\end{figure*}

SSA is a supervised learning technique where an orthogonal transformation matrix is learned for every college in the postsecondary system. This orthogonal matrix works like an encoder to a vector space shared across colleges while its transpose--which is also its inverse--decodes back to the college’s local vector space (Fig. \ref{fig:fig4}). We formulate the problem as the following \cite{chen2018unsupervised,artetxe2017unsupervised,lample2017unsupervised}:

\begin{equation}
M^* = \arg\min_{M} \, \lVert x_i M_i M_j^\top - x_j \rVert \quad \forall \, i,j \in U,
\end{equation}
\[
\begin{aligned}
\text{where } 
&M: \text{ set of transformation matrices}, \\
&M_i \in \mathbb{R}^{d \times d}, \quad M_j^{-1} = M_j^\top, \\
&U: \text{ set of all colleges within the postsecondary system}, \\
&x_i, x_j: \text{ vector representations of equivalent courses arranged in corresponding order.}
\end{aligned}
\]

Given the absence of an analytical solution \cite{chen2018unsupervised}, we use supervised learning to arrive at $M^*$. We iterate through pairs of colleges to minimize the error between the predicted location of a course in another college’s vector space and the vector representation of the true equivalent course from existing agreements.

% FIG 4

When training SSA and evaluating its performance, we used a 5-fold cross-validation strategy. The articulation data was divided into five equal parts; in each fold, four-fifths of the data were used for supervised training and the remaining one-fifth for evaluation. Model performance was computed by aggregating the number of correct predictions across all folds and dividing by the total number of evaluation examples.

\subsection{Evaluation}
Our evaluation examines two dimensions: (1) whether SSA more accurately predicts the articulation relationships defined by human experts, and (2) whether it better captures academic and disciplinary proximity rather than surface text similarity. For (1), we compared the performance of different vectorization methods across three generations of NLP models, both with and without SSA, using recall against faculty-established equivalencies. For (2), we measured changes in the dispersion of course embeddings within Classification of Instructional Programs (CIP) \footnote{\url{https://nces.ed.gov/ipeds/cipcode/Default.aspx?y=56}} categories using the effective radius metric, which computes the average Euclidean distance of embeddings from their category centroid.

We began by evaluating how accurately algorithmic models predict existing faculty-established articulations by experimenting with three generations of NLP models for vectorization and comparing their performance across generations: the first generation, Word2Vec, which introduced neural network-based word embeddings \cite{fix1985discriminatory} and is referred to here as DescVec; the second generation, SBERT, the first widely adopted transformer-based model for comprehensive natural language process \cite{reimers2019sentence}; and the third generation, GPT-style class of large language models (LLMs) that incorporate advancements in alignment techniques and Reinforcement Learning from Human Feedback (RLHF \cite{lloyd1982least}), which we refer to as OpenAI. We also experimented with integrating the sequence of course IDs from student enrollment histories (i.e., Course2vec) with pairwise alignments \cite{pardos2019data} and SSA into these NLP models, hypothesizing that this enrollment data could add valuable signal to a course representation \cite{pardos2020university}. Each model was tested with and without SSA, trained using an existing subset of articulation agreements. To evaluate these vectorization methods, we applied KNN to predict candidate articulations and assessed performance using recall@1 and recall@5 metrics, based on faculty-established articulations.

We then conducted evaluations of the effects of SSA to examine whether it effectively positions courses based on their underlying academic content. To assess this, we compared the dispersion of course embeddings within each CIP category using OpenAI embeddings generated with and without SSA. CIP is a taxonomy developed by the U.S. Department of Education’s National Center for Education Statistics to classify postsecondary instructional programs and provides a nationally recognized, academic content-based reference system. As a federally maintained taxonomy used for reporting and policy across U.S. higher education, CIP serves as a reliable external benchmark for testing whether embeddings capture disciplinary similarity rather than superficial catalog language.

For each CIP category $p$, we quantify the dispersion of course vectors using an effective radius $r_g^{(p)}$, analogous to the radius of gyration. Let $c_i^{(p)}$ denote the embedding vector of the $i$-th course in CIP $p$, $\bar{c}^{(p)}$ the centroid of all $N_p$ courses in that CIP, and $\|\cdot\|$ the Euclidean norm. The effective radius is defined as

\[
r_g^{(p)} = \sqrt{\frac{1}{N_p} \sum_{i=1}^{N_p} \|c_i^{(p)} - \bar{c}^{(p)}\|^2}.
\]

This metric captures the average distance of courses from their CIP centroid; a smaller $r_g^{(p)}$ indicates that courses sharing a CIP code are more tightly clustered in the embedding space.

We also conducted qualitative analyses to examine how SSA mitigates non-content-based grouping. Specifically, we compared course visualizations with two sets of OpenAI embeddings generated—with and without SSA. We then applied t-distributed stochastic neighbor embedding (t-SNE) for dimensionality reduction and visualized the two sets separately. In the visualizations, each node represents a course and is colored by its CIP code. We then compared the two plots to identify apparent clusters that showed the most noticeable shifts after applying SSA and examined the corresponding course titles and catalog descriptions to determine whether linguistic or institutional artifacts (e.g., repeated disclaimers or shared boilerplate) were responsible for the spurious grouping observed prior to SSA.

\subsection{Articulation Expansion using Similarity Threshold} 
Using the aligned embeddings, we further assessed how many additional course equivalencies could be identified, setting a threshold for course similarity based on the Area Under the Receiver Operating Characteristic Curve (AUC-ROC). 

SSA is built upon MRSR \cite{chen2018unsupervised}, yet a key limitation of word-to-word language translation work is the underlying assumption that an equivalent pair always exists, which leads the model to return the most similar translation (nearest neighbor) even when no true equivalent is available. This assumption does not hold in our context of higher stakes course articulations, where not all courses have valid counterparts across institutions. To address this, we use negative sampling and AUC-ROC to establish a similarity threshold using existing articulation agreements, enabling the identification of previously undetected course equivalencies.

\begin{figure}[!htbp]
\centering
\includegraphics[width=1\textwidth]{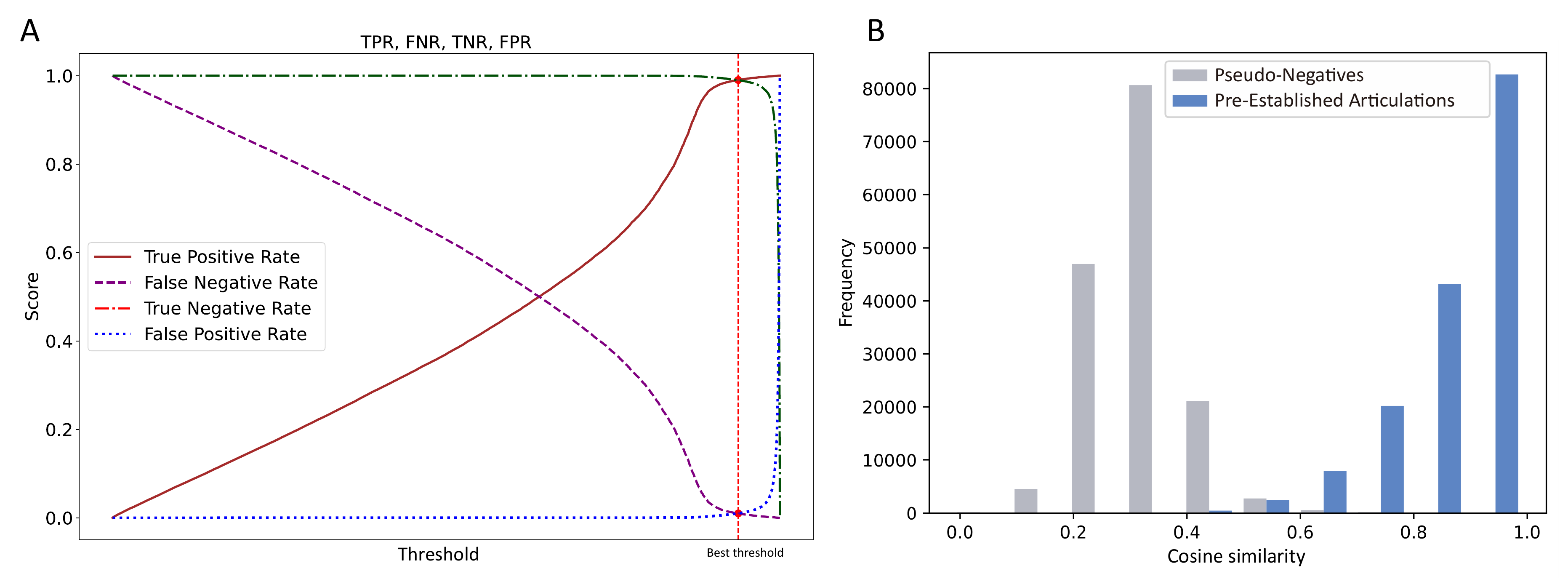}
\caption{\textbf{An example of similarity thresholding for course articulation expansion, based on SSA OpenAI embeddings for SUNY courses.} A. Relationship between threshold values and TPR, FNR, TNR, and FPR. The optimal threshold is selected to balance maximizing the True Positive Rate (TPR) while minimizing the False Positive Rate (FPR). Note that TPR = 1 - FNR and TNR = 1 - FPR. B. Distribution of cosine similarities between course pairs: pseudo-negatives vs. pre-established articulations. Existing articulation pairs have higher cosine similarities (average 0.869), while pseudo-negative pairs have lower similarities (average 0.321).}
\label{fig:fig5}
\end{figure}

Deciding if a pair of courses qualify as a valid equivalency is akin to a binary classification task. Leveraging machine learning techniques in binary classification, such as optimizing AUC-ROC, we determine the optimal threshold for the cosine distance between courses that maximizes their likelihood of being equivalent. From this analysis, we determine the threshold that balances maximizing True Positive Rate (TPR) while minimizing False Positive Rate (FPR). The chosen threshold, referred to as the best threshold in Fig. \ref{fig:fig5}A, is the cosine similarity value above which courses are classified as equivalent.

Unlike classical binary classification, we are limited by the lack of true negatives. Since the absence of a pre-existing articulation agreement does not necessarily imply that two courses are truly non-equivalent, we generate a set of \textit{pseudo-negatives} by randomly sampling course pairs that do not have an existing articulation record. In SUNY, our pseudo-negatives consist of same-level lower division courses, bringing their average distance closer to that of the true articulations which consist mostly of lower division courses. This results in a more conservative threshold, as the ROC curve shifts toward higher cosine similarity values. Fig. \ref{fig:fig5}B demonstrate that the juxtaposed histograms of the cosine similarities of pseudo-negatives and true positives (i.e. equivalent course pairs from the ground truth) can be seen separated around a cosine similarity threshold. The AUC-ROC analysis of the combined set of true articulations and pseudo-negative samples yields this optimal threshold. As a result, the threshold for SUNY is 0.547. 

To identify potential articulations that may have been undetected by human experts, we used KNN to predict the top candidate articulations above a specific threshold. Knowing the cosine similarity threshold in a vector space allows us to identify equivalent course pairs with high confidence. We perform a combinatorial search to find course pairs whose vectors exceed this similarity threshold. For SUNY, one course typically articulates to a single course at the destination institution ($\mathit{m} = 1.050$, $\sigma = 0.236$). Accordingly, we identify the top-1 most cosine-similar course for each source course. To focus on newly identified equivalencies, we discard course pairs already present in the ground truth and report only those additionally retrieved through the nearest neighbor search in vector space.

%fig5 here

\section{Results}

\subsection{Course-to-Course Articulation Prediction}
\label{result-prediction}
\begin{figure}[!htbp]
\centering
\includegraphics[width=0.9\textwidth]{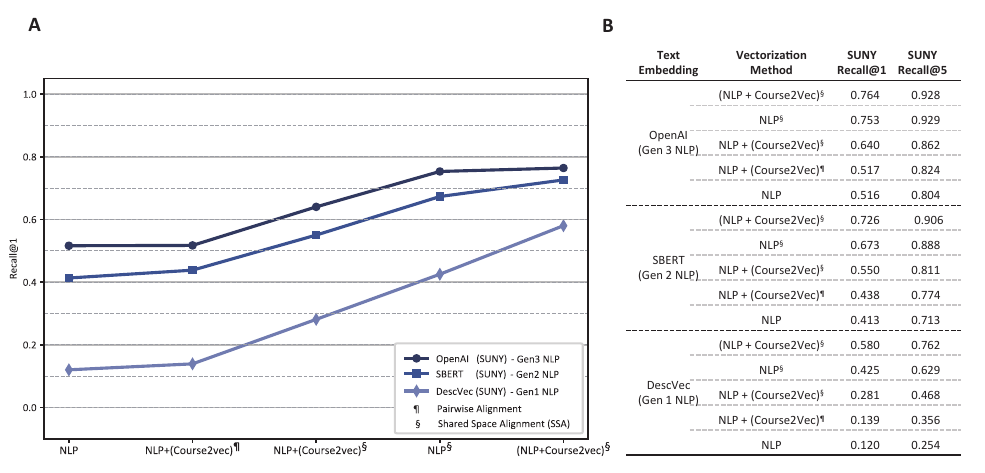}
\caption{\textbf{Course-to-course articulation prediction results.} A. Recall@1 scores of vectorization methods across different NLP generations for SUNY. + denotes concatenation of course vectors. B. Recall@1 and recall@5 scores. Notations are the same as A.}
\label{fig:fig6}
\end{figure}

\begin{figure*}[!htbp]
\centering
\includegraphics[width=0.8\textwidth]{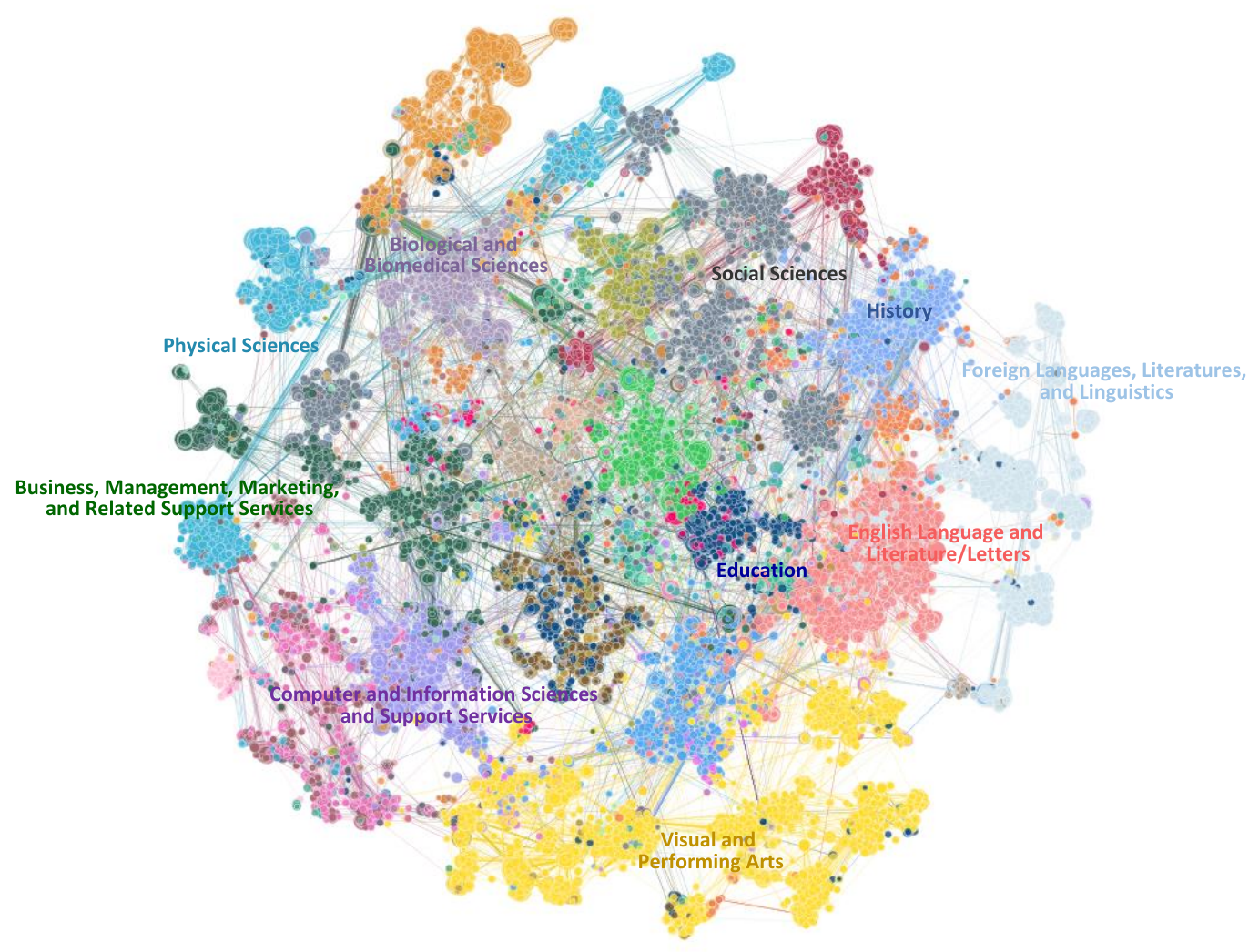}
\caption{\textbf{SUNY course articulation map.} t-SNE visualization of courses generated using the OpenAI+SSA model. Courses are represented as nodes, with the size of each node corresponding to the number of courses it is articulated with. Edges between nodes indicate current course articulation agreements between different institutions.The annotations represent the two-digit Classification of Instructional Programs (CIP) codes for the top 10 fields with the most courses.}
\label{fig:fig3}
\end{figure*}

%Our best method achieved a recall@1 of 0.764, compared to 0.139 with the best method from previous research~\cite{pardos2019data}.
Course articulation within the SUNY system is intended to identify courses across its campuses that are sufficiently similar to be considered transferable between a pair of institutions. Figures \ref{fig:fig6}A and \ref{fig:fig6}B show the prediction performance results for five different models across three generations of NLP, highlighting that performance improves with each subsequent generation. The average recall@1 score for all models is 0.309 for the first generation, 0.560 for the second, and 0.638 for the third. The best-performing method from previous research~\cite{pardos2019data} achieved a recall@1 of 0.139 and a recall@5 of 0.356. Overall, the combination of the OpenAI and Course2vec models with SSA achieves the highest performance, with recall@1 reaching 0.764 and recall@5 achieving 0.928. The primary performance boost comes from SSA, which increases performance by an average of 121.017\% from NLP alone to NLP with SSA across all generations. The integration of Course2vec with SSA adds an additional 15.269\% improvement on average. However, as the NLP models advance through generations, the relative impact of Course2vec on performance diminishes. To illustrate the articulation structure, Fig. \ref{fig:fig3} provides a visualization of existing SUNY articulation agreements using OpenAI + SSA model.

%fig 6

% articulation map

\subsection{Effects of SSA on CIP Dispersion}
\label{sec:ssa-effects}
\begin{figure*}[!htbp]
\centering
\adjustbox{max width=\textwidth, max totalheight=\textheight}{%
    \includegraphics{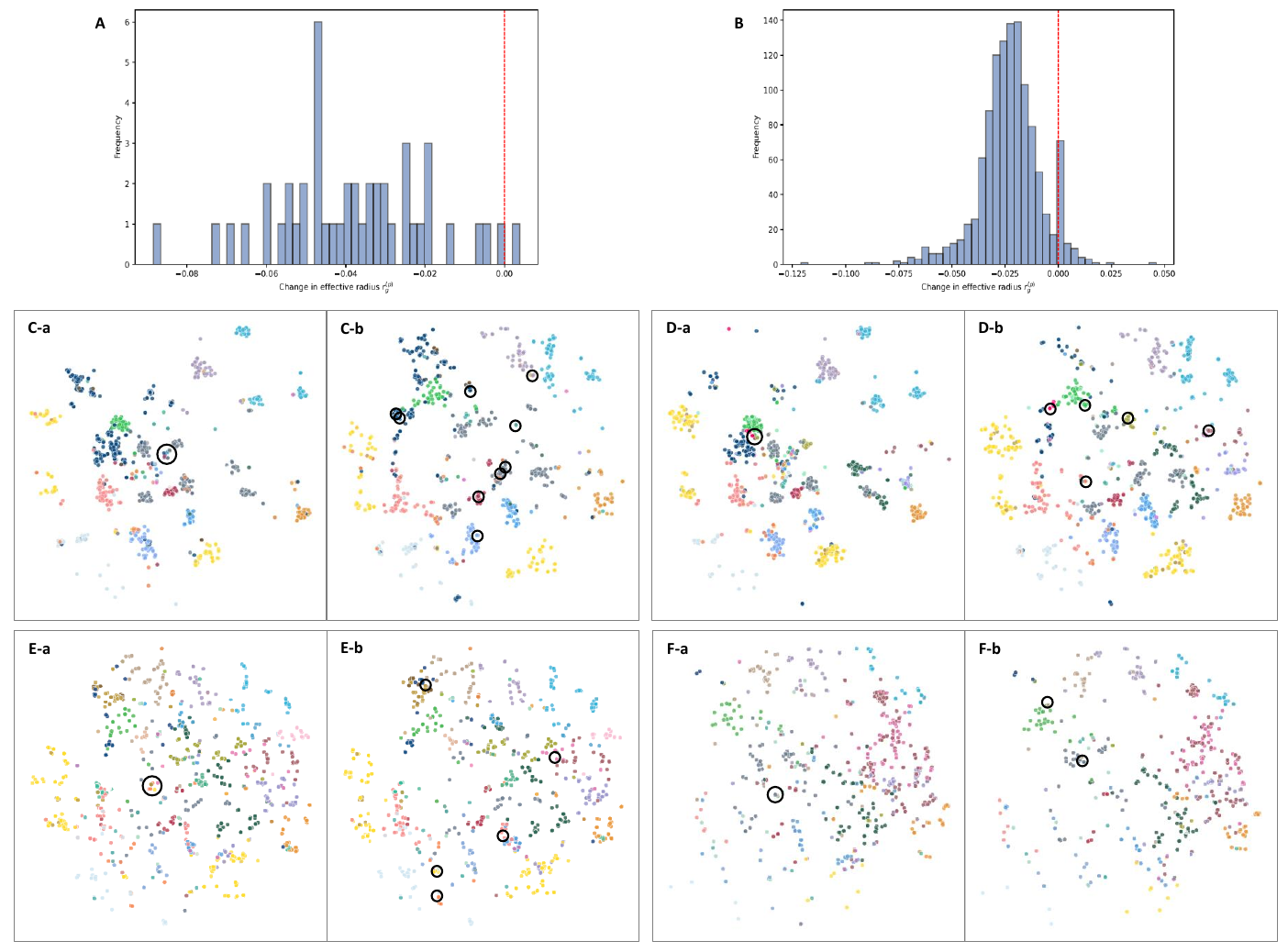}
}
\caption{\textbf{Quantitative and qualitative analyses of SSA effects.} (A-B) Distributions of changes in effective radius across CIP codes, with the red dashed line indicating zero. Bars below the line represent CIP categories where dispersion decreased after applying SSA. Panel A represents system-wide CIP dispersion, while panel B shows institutional-level CIP dispersion. (C-F) t-SNE visualizations comparing course embeddings before (-a) and after (-b) applying SSA. Each dot represents a course, colored by CIP code, and circles highlight clusters that exhibited notable shifts toward content-based alignment.}
\label{fig:fig7}
\end{figure*}

% maybe appendix needed, esp qual analysis
We examined changes in $r_g^{(p)}$ at two levels: system-wide CIP radius and institutional-level CIP radius. Across all campuses, SSA reduced dispersion for 45 of the 46 CIP categories (Fig. \ref{fig:fig7}A), with a mean difference of -0.038 compared to embeddings without SSA. This indicates that, after alignment, courses sharing the same CIP are positioned more closely across institutions, demonstrating that SSA aligns cross-campus equivalents according to shared academic content.

Within each individual college, we also measured $r_g^{(p)}$ for each CIP using only the courses of that college. SSA reduced dispersion for 91.76\% of the institution-CIP pairs (Fig. \ref{fig:fig7}B), with a mean difference of -0.023, indicating that SSA positions courses according to their academic content, even though repeated generic terms and textual artifacts may exist within campuses.

From qualitative analysis, we observed that before applying SSA, 11 internship courses from different departments at SUNY Cortland were closely clustered (Fig. \ref{fig:fig7}C-a). These courses span six distinct CIP codes, yet all include the term “\textit{Internship},” in their titles, such as \textit{Internship in Public History} or \textit{Pre-Law Internship}. Their descriptions are abbreviated with department codes like HIS or POL, without further details. Given this limited textual information, the language model likely grouped these courses based on the common term “\textit{Internship}.” However, SSA appeared to capture content signals beyond the generic title, reducing the influence of this institution-specific repetition and repositioning each course closer to others sharing the same CIP classification (Fig. \ref{fig:fig7}C-b).

Figure \ref{fig:fig7}D shows eight courses from SUNY College in Oswego with five different CIP codes. These eight courses are all practicum courses from various departments. Each course title includes the word “\textit{Practicum},” such as \textit{Practicum in Teaching} or \textit{Practicum in Public Justice}. After SSA, each moved closer to courses within its correct CIP.

Figure \ref{fig:fig7}E shows seven courses from Suffolk County Community College spanning six CIPs. Each course title shares the word “\textit{Independent study},” such as \textit{Independent Study: Chinese} or \textit{Independent Study: Fitness Specialist.} After applying SSA, the influence of this overlapping term was reduced, positioning each course closer to others within the same CIP classification.

Figure \ref{fig:fig7}F shows six courses from SUNY Polytechnic Institute that formed a tight cluster despite spanning three distinct CIP codes. This clustering likely occurred because each catalog description includes the same institutional disclaimer—“\textit{Students may receive credit in a future semester for different topic areas}”—and the titles share the word “\textit{Topics},” such as \textit{Special Topics in Psychology} or \textit{Special Topics in Sociology}. SSA reduced this grouping, dispersing the courses toward neighborhoods consistent with their respective CIP classifications.

\subsection{Articulation Expansion}
\label{result-articulation-expansion}
Using OpenAI with SSA, we identified 2,787,526 additional above-similarity-threshold articulation pairs, which is 17.759 times greater than existing agreements. Given that the SUNY system is multisegmental, comprising both 2-year community colleges and 4-year universities, we further analyzed the results by institution type. The breakdown revealed a substantial increase in potential articulation pairs across all institution types: 245,326 pairs for 2-year to 2-year colleges (13.07 times more than existing agreements), 324,431 pairs for 2-year to 4-year colleges (5.502 times more), 1,305,605 pairs for 4-year to 4-year colleges (24.327 times more), and 912,164 pairs for 4-year to 2-year colleges (35.671 times more).

\subsection{Projected Adoption and Impact of Expanded Articulations}
The average adoption rate across the two surveys was 61.23\%, with staff and faculty showing comparable rates. In Section \ref{result-articulation-expansion}, we reported that our method identifies an additional 2,787,526 potential articulations. If we assume that the stakeholder preferences observed in our survey would hold, we can apply the average committee adoption rate of 61.23\% to these newly identified pairs. Under this scenario, approximately 1,706,802 of these pairs could be accepted by faculty and staff. Compared to the 156,968 existing articulations, this projection represents an 11.87-fold increase in valid credit-mobility opportunities.

% This projection is supported in two ways. -> lower bound
This projection as a conservative estimate is supported in two ways. First, the improved accuracy of our novel method increases the likelihood that faculty and staff would adopt its recommendations. Figure \ref{fig:fig6} shows that SBERT achieves a recall@1 of 0.413 and recall@5 of 0.713, whereas our best method achieves 0.764 and 0.928. Higher accuracy reduces the risk of presenting stakeholders with irrelevant or incorrect recommendations, which prior work shows is an important determinant of expert acceptance~\cite{xu2023convincing,highhouse2008stubborn}. Second, the main concerns expressed by faculty and staff--the need for semantically and academically accurate, rather than lexically superficial, recommendations--are addressed by SSA. This is reflected in the tightening of CIP-based clusters shown in Section \ref{sec:ssa-effects} and Figure~\ref{fig:fig7}. These results suggest that the adoption patterns observed in the survey provide a reasonable lower bound to estimate the adoption of the articulations generated by our improved model. 

\section{Discussion}

%summarizes the survey + discussion (percentage of adoption ---higher than the field expects (higher ed)- cite (overly critical faculty evaluation, faculty insistent on keeping authority )

%In the two surveys, the overall adoption rate averaged 61.23\%, with staff adopting 63.25\% of recommendations and faculty 59.21\%, indicating comparable levels of acceptance. Prior work shows that people are prone to abandon algorithmic recommendations after observing an error \cite{dietvorst2015algorithm}, and this tendency is likely to persist in higher education, where professional judgment is highly valued and change is often constrained by loss aversion and a preference for maintaining existing practices \cite{tagg2012does}. Against this backdrop, an adoption rate of 61.23\% is higher than might reasonably be anticipated. One possible interpretation is that course articulation work, which is labor-intensive and requires reviewing large volumes of information, may make stakeholders perceive AI assistance as reducing cognitive load rather than threatening their expertise. In line with studies showing that users respond more favorably to AI when it offers clear resource or efficiency gains \cite{bankins2024multilevel}, such perceived reductions in cognitive load may help explain the observed willingness to adopt AI recommendations even in a traditionally change-resistant domain.

In the two surveys, the overall adoption rate averaged 61.23\%, with staff adopting 63.25\% of recommendations and faculty 59.21\%, indicating comparable levels of acceptance. This adoption rate is likely higher than expected by practitioners, given the barriers to articulation perceived by the field, particularly with respect to faculty \cite{renaud2000examination}. One possible interpretation is that course articulation work, which is labor-intensive and requires reviewing large volumes of information, may make stakeholders perceive AI assistance as reducing cognitive load rather than threatening their expertise. In line with studies showing that users respond more favorably to AI when it offers clear resource or efficiency gains \cite{bankins2024multilevel}, such perceived reductions in cognitive load may help explain the observed willingness to adopt AI recommendations even in a traditionally change-resistant domain.
%prior work suggesting that faculty hesistant to adopt articulation - overly critical/cautious 
% not many approvals (articulation context) 
%interviews with faculty (default reject)

%The barriers from the 2 1 themes, which emerged from question 1 ; barriers, which are of a perceptual nature, rather than factual, comprise most of the top 5 ranked themes, and include: elitist attitudes (ranked 1); quality debate - diflerent curriculum (ranked 2); history (ranked 3 - tie); will of people involved (tanked 5); lack of understanding of each other - values (ranked 6); fear of encroachment (ranked 7); credentials of college faculty (ranked 10); misperception of what articulation is (ranked 12); colleges themselves (ranked 14); and lack of data on success (ranked 16). 

Improving accuracy is essential for gaining the trust of human experts, as the presence of errors in AI suggestions often lead to the rejection of recommendations due to the high expectations placed on AI~\cite{xu2023convincing,highhouse2008stubborn}. Our best method achieved a recall@1 of 0.764, compared to 0.139 with the best method from previous research~\cite{pardos2019data}. Applying SSA alone to the same baseline raised recall@1 from 0.139 to 0.281, representing a 2.02 times improvement over the prior best method~\cite{pardos2019data}. Overall, the primary performance gain comes from incorporating SSA, resulting in a 62.95\% improvement in the second-generation model and a 45.9\% improvement in the third-generation model. Furthermore, our results also show that valuable signals can also be found in student enrollment histories, consistent with previous studies \cite{pardos2020university}. Incorporating Course2vec improved accuracy, with the largest gains in the first generation of NLP models, followed by the second and third. This is likely because earlier NLP models were less capable of capturing all relevant information from catalog descriptions, making enrollment data a useful complement. Overall, adding Course2vec yielded an average 15.27\% increase in recall@1 across all NLP generations. 

By incorporating stakeholder input into its design, SSA aligns with faculty and staff expectations and directly addresses their concerns about algorithmic recommendations. Our findings show that SSA effectively reduces the influence of institution-specific artifacts and surface-level lexical overlap, thereby enabling more robust, content-based similarity detection. Consequently, courses are positioned more accurately according to their disciplinary content, reflected in the greater within-CIP consistency of the embedding space. Prior research suggests that AI systems attuned to human preferences and perceptions can improve decision-making outcomes~\cite{russell2019human,burton2020systematic} and foster trust in AI-assisted processes~\cite{burton2020systematic}. Such trust, in turn, can enhance users’ perceptions of an AI system’s performance~\cite{pataranutaporn2023influencing}. In this sense, SSA not only enhances the technical accuracy of course-articulation recommendations but also supports more transparent and collaborative decision-making. These qualities may ultimately increase faculty willingness to engage with and adopt AI-assisted articulation systems.

Using a thresholding method based on faculty-established articulations, we identified 17.759 times more equivalent course pairs, suggesting an additional 2,787,526 potential articulations. With an average stakeholder acceptance rate of 61.23\%, this suggests 1,706,802 additional credit mobility opportunities that would otherwise have gone undetected. Such under-recognition of valid equivalencies may result in credit loss for students. Studies found that losing more than seven credits during transfer negates the cost advantage of starting at a community college over a four-year institution in certain states \cite{belfield2017really}. 

%C-RAC crac2025ai
This study suggests that AI can streamline the articulation process, indicating a potential shift in current practice. While final approval remains a policy-driven, socio-technical task requiring faculty and staff oversight \cite{ntftac2021reimagining}, our approach narrows the pool of candidate courses, surfacing 92\% of articulation matches within the top five suggestions. This capability could help transition the current practice, which follows a policy requiring manual approval for every pair of equivalencies, to one where a committee of faculty audits a sample of the algorithm’s predicted equivalencies and provides feedback to maintain alignment with institutional standards. Institutions are showing increasing interest in this potential shift, as reflected in the recent statement by the Council of Regional Accrediting Commissions encouraging the use of AI to streamline workflows and identify new course equivalencies \cite{crac2025ai}.

Maximizing the benefits of expanded course equivalencies requires increasing student awareness. While these agreements are valuable tools for establishing transfer pathways, they are ineffective if not successfully shared with students \cite{logue2024possible,banuelos2024community}. Academic advising is crucial in this effort, as research shows advisors play a critical role in helping students navigate the transfer process and overcome barriers \cite{hayes2020role}. With the support of this infrastructure, advisors can better assist students who are uncertain about their transfer destination or intended major. However, given the high advisor-to-student ratios at community colleges, often reaching as high as 1:1,200 \cite{mayer2019integrating}, AI-augmented advising systems could provide support by methodically evaluating transfer requirements at a variety of destinations \cite{nguyen2025community}. By combining AI-assisted course articulations with improved access to transfer pathways, institutions could better support students toward their intended transfer outcomes.

% credit mobility - we can apply to other related contexts (Discussion/future work) applicable to ~~ credit equivalency scenarios - ccn, high school to university, cpl (credit for prior learning/ prior learning assessment)
Beyond articulation, this approach could also be applied to other higher education contexts where identifying equivalencies or alignments among courses or learning outcomes is essential. For example, it could support initiatives such as Common Course Numbering, which uses a uniform numbering system for comparable lower-division courses across postsecondary institutions to facilitate transfer--an approach already adopted in many states such as Arizona and California~\cite{californiaAB1111}. Similarly, it could be leveraged in Credit for Prior Learning (CPL) processes to help institutions evaluate how students’ prior learning experiences align with existing course requirements~\cite{ryu2013credit}. Applying this approach to these broader contexts could help institutions better recognize students’ diverse educational pathways.

% revised
This study has two limitations that could suggest directions for future work. Our adoption projections were based on responses from real faculty and staff; however, they were not based on recommendations from our improved accuracy model and were considered in a lower-stakes context where stakeholders expressed an intent-to-adopt as opposed to finalizing articulation adoption that would affect students. It is not clear if an expected increase in adoption rate due to improved model accuracy would be greater or less than the decrease in adoption expected due to increased scrutiny in a higher-stakes scenario. Future research could investigate the adoption of articulations generated with our improved method in real review settings to more accurately capture stakeholder behavior. Second, although SSA significantly improved performance, it relies on existing articulation pairs, making it impractical to apply in institutions where no course articulation data is available. This limitation underscores the need for future approaches that can reduce dependency on human-provided data to conduct alignment.

As transfer demands grow \cite{nscrc2024transfer}, so too will the need for large-scale adoption of AI-assistive technologies and policies. Ensuring that these technologies are designed with stakeholder input will be critical to improving student transfer pathways and advancing equity in higher education.

%%%%%%%%%%%%%%%%%%%%%%%%%%%%%%%%%%%%%%%%%%%%%%%%%%%%%%%%%%%%%%%%%%%%%%%%%%%%%%%%%%%%%%%%%%%%%%%%%%%%%%%%%%%%%%%%%%%%%%%%%%%%%%%%%%%%%%%%%%%%%%%%%%%%%%%%%%%%%%%%%%%%%%%%%%

\section*{Acknowledgments}

This research was supported by a grant from the Gates Foundation (Award \#91974). Our dataset for this work was provisioned, in part, by the SUNY Office of Institutional Research. We thank Thomas Hanford, Assistant Vice Chancellor for Transfer and Articulation at the State University of New York, and Dan Knox, Director, NASH Institute for Systems Innovation \& Improvement, for insights and additional data support. We thank faculty from SUNY and institutional staff from the New York State Transfer and Articulation Association for participating in our adoption surveys. Finally, we thank Jenny Jiang for her early contributions to the similarity thresholding approach, Yueqi Wang for early contributions to the articulation evaluation codebase, and Shreya Sheel for her work deploying the faculty survey. 

The views expressed in this research are those of the authors and do not necessarily reflect the positions of the Gates Foundation, SUNY, or other individuals acknowledged herein. This research was approved by the UC Berkeley Committee for the Protection of Human Subjects under IRB Protocol 2022-04-15286.

%Bibliography
\bibliographystyle{unsrt}  
\bibliography{references}

@article{dietvorst2015algorithm,
  title={Algorithm aversion: people erroneously avoid algorithms after seeing them err.},
  author={Dietvorst, Berkeley J and Simmons, Joseph P and Massey, Cade},
  journal={Journal of experimental psychology: General},
  volume={144},
  number={1},
  pages={114},
  year={2015},
  publisher={American Psychological Association}
}

@book{filiz2021tragedy,
  title={The tragedy of algorithm aversion},
  author={Filiz, Ibrahim and Judek, Jan Ren{\'e} and Lorenz, Marco and Spiwoks, Markus},
  year={2021},
  publisher={Ostfalia Hochschule f{\"u}r Angewandte Wissenschaften, Fakult{\"a}t Wirtschaft}
}

@article{burton2020systematic,
  title={A systematic review of algorithm aversion in augmented decision making},
  author={Burton, Jason W and Stein, Mari-Klara and Jensen, Tina Blegind},
  journal={Journal of behavioral decision making},
  volume={33},
  number={2},
  pages={220--239},
  year={2020},
  publisher={Wiley Online Library}
}

@article{bohlen2025overcoming,
  title={Overcoming Algorithm Aversion with Transparency: Can Transparent Predictions Change User Behavior?},
  author={Bohlen, Lasse and Kruschel, Sven and Rosenberger, Julian and Zschech, Patrick and Kraus, Mathias},
  journal={arXiv preprint arXiv:2508.03168},
  year={2025}
}

@article{velasco2024tracking,
  title={Tracking Transfer: Four-Year Institutional Effectiveness in Broadening Bachelor's Degree Attainment.},
  author={Velasco, Tatiana and Fink, John and Bedoya, Mariel and Jenkins, Davis and LaViolet, Tania},
  journal={Community College Research Center, Teachers College, Columbia University},
  year={2024},
  publisher={ERIC}
}

@article{belfield2017really,
  title={Is it really cheaper to start at a community college? The consequences of inefficient transfer for community college students seeking bachelor’s degrees},
  author={Belfield, Clive and Fink, John and Jenkins, Paul Davis},
  journal={Community College Research Center, Teachers College, Columbia University},
  year={2017}
}

@book{bender1990spotlight,
  title={Spotlight on the Transfer Function: A National Study of State Policies and Practices.},
  author={Bender, Louis W},
  year={1990},
  publisher={American Association of Community and Junior Colleges}
}

@book{donovan1987transfer,
  title={Transfer: Making It Work. A Community College Report.},
  author={Donovan, Richard A et al.},
  year={1987},
  publisher={American Association of Community and Junior Colleges, National Center for Higher Education}
}

@book{knoell1990transfer,
  title={Transfer, Articulation, and Collaboration: Twenty-Five Years Later.},
  author={Knoell, Dorothy},
  year={1990},
  publisher={American 	Association of Community and Junior Colleges}
}

@article{boggs2010growing,
  title={Growing roles for science education in community colleges},
  author={Boggs, George R},
  journal={Science},
  volume={329},
  number={5996},
  pages={1151--1152},
  year={2010},
  publisher={American Association for the Advancement of Science}
}

@article{hodara2016improving,
  title={Improving credit mobility for community college transfer students: Findings and recommendations from a 10-state study},
  author={Hodara, Michelle and Martinez-Wenzl, Mary and Stevens, David and Mazzeo, Christopher},
  journal={Planning for Higher Education},
  volume={45},
  number={1},
  pages={50--118},
  year={2016},
  publisher={Society for College and University Planning}
}

@article{monaghan2015community,
  title={The community college route to the bachelor’s degree},
  author={Monaghan, David B and Attewell, Paul},
  journal={Educational Evaluation and policy analysis},
  volume={37},
  number={1},
  pages={70--91},
  year={2015},
  publisher={Sage Publications Sage CA: Los Angeles, CA}
}

@article{hodara2017exploring,
  title={Exploring credit mobility and major-specific pathways: A policy analysis and student perspective on community college to university transfer},
  author={Hodara, Michelle and Martinez-Wenzl, Mary and Stevens, David and Mazzeo, Christopher},
  journal={Community College Review},
  volume={45},
  number={4},
  pages={331--349},
  year={2017},
  publisher={SAGE Publications Sage CA: Los Angeles, CA}
}

@misc{us2017higher,
  title={Higher education: Students need more information to help reduce challenges in transferring college credits},
  author={US Government Accountability Office},
  year={2017},
  publisher={Author Washington, DC}
}

@article{logue2024possible,
  title={Possible causes of leaks in the transfer pipeline: Student views at the 19 colleges of The City University of New York},
  author={Logue, AW and Oka, Yoshiko and Wutchiett, David and Gentsch, Kerstin and Abbeyquaye, Stephanie},
  journal={Journal of College Student Retention: Research, Theory \& Practice},
  volume={26},
  number={3},
  pages={721--750},
  year={2024},
  publisher={SAGE Publications Sage CA: Los Angeles, CA}
}

@inproceedings{pardos2019data,
  title={Data-assistive course-to-course articulation using machine translation},
  author={Pardos, Zachary A and Chau, Hung and Zhao, Haocheng},
  booktitle={Proceedings of the Sixth (2019) ACM Conference on Learning@ Scale},
  pages={1--10},
  year={2019}
}

@article{kizilcec2023pipelines,
  title={From pipelines to pathways in the study of academic progress},
  author={Kizilcec, Ren{\'e} F and Baker, Rachel B and Bruch, Elizabeth and Cortes, Kalena E and Hamilton, Laura T and Lang, David Nathan and Pardos, Zachary A and Thompson, Marissa E and Stevens, Mitchell L},
  journal={Science},
  volume={380},
  number={6643},
  pages={344--347},
  year={2023},
  publisher={American Association for the Advancement of Science}
}

@misc{ecs2022transfer,
  author = {{Education Commission of the States}},
  title = {Transfer and Articulation Policies: 50-State Comparison},
  year = {2022},
  url = {https://reports.ecs.org/comparisons/transfer-and-articulation-2022},
  note = {Accessed: 2025-05-05}
}

@article{ignash2000evaluating,
  title={Evaluating state-level articulation agreements according to good practice},
  author={Ignash, Jan M and Townsend, Barbara K},
  journal={Community College Review},
  volume={28},
  number={3},
  pages={1--21},
  year={2000},
  publisher={Sage Publications Sage CA: Thousand Oaks, CA}
}

@book{ciac2013handbook,
  author = {{California Intersegmental Articulation Council}},
  title = {California Articulation Policies and Procedures Handbook},
  year = {2013},
  publisher = {California Intersegmental Articulation Council}
}

@inproceedings{xu2023convincing,
  title={Convincing the Expert: Reducing Algorithm Aversion in Administrative Higher Education Decision-making},
  author={Xu, Lingrui and Pardos, Zachary A and Pai, Anirudh},
  booktitle={Proceedings of the Tenth ACM Conference on Learning@ Scale},
  pages={215--225},
  year={2023}
}

@article{barry1992establishing,
  title={Establishing equality in the articulation process},
  author={Barry, Roger J and Barry, Phyllis A},
  journal={New directions for community colleges},
  volume={78},
  number={2},
  pages={35--44},
  year={1992},
  publisher={Citeseer}
}

@techreport{katsinas2016alabama,
  title = {Alabama Articulation and General Studies Committee \& Statewide Transfer and Articulation Reporting System: Evaluation Project Final Report},
  author = {Katsinas, Stephen G and Bray, Nathaniel P and Dotherow, James E and Malley, Michael S and Warner, Jake L and Adair, J Lucas and Roberts, John and Phillips, Undre V},
  year = {2016},
  institution = {Education Policy Center}
}

@misc{suny2020history,
  author = {{The State University of New York}},
  title = {History of SUNY},
  year = {2020},
  url = {https://www.suny.edu/about/history/},
  note = {Accessed: 2025-05-05}
}

@book{fix1985discriminatory,
  title={Discriminatory analysis: nonparametric discrimination, consistency properties},
  author={Fix, Evelyn},
  volume={1},
  year={1985},
  publisher={USAF school of Aviation Medicine}
}

@techreport{ntftac2021reimagining,
  author = {{National Task Force on the Transfer and Award of Credit}},
  title = {Reimagining Transfer for Student Success},
  year = {2021},
  institution = {American Council on Education}
}

@article{highhouse2008stubborn,
  title={Stubborn reliance on intuition and subjectivity in employee selection},
  author={Highhouse, Scott},
  journal={Industrial and organizational psychology},
  volume={1},
  number={3},
  pages={333--342},
  year={2008},
  publisher={Cambridge University Press}
}

@article{pardos2020university,
  title={A university map of course knowledge},
  author={Pardos, Zachary A and Nam, Andrew Joo Hun},
  journal={PloS one},
  volume={15},
  number={9},
  pages={e0233207},
  year={2020},
  publisher={Public Library of Science San Francisco, CA USA}
}

@article{banuelos2024community,
  title={Community College Counselors’ Experiences Assisting Community College Students in Navigating Course-Taking at Multiple Institutions},
  author={Ba{\~n}uelos, Maricela and Flores, Kassandra and Rozhenkova, Veronika and Morales-Gracia, Maritza and Cabrera, Jennifer and Baker, Rachel},
  journal={Community College Journal of Research and Practice},
  pages={1--16},
  year={2024},
  publisher={Taylor \& Francis}
}

@article{hayes2020role,
  title={The role of academic advisors in the development of transfer student capital},
  author={Hayes, Shannon and Lindeman, Leslie and Lukszo, Casey},
  journal={NACADA Journal},
  volume={40},
  number={1},
  pages={49--63},
  year={2020},
  publisher={The Global Community for Academic Advising}
}

@article{mayer2019integrating,
  title={Integrating Technology and Advising: Studying Enhancements to Colleges' iPASS Practices.},
  author={Mayer, Alexander and Kalamkarian, Hoori Santikian and Cohen, Benjamin and Pellegrino, Lauren and Boynton, Melissa and Yang, Edith},
  journal={MDRC},
  year={2019},
  publisher={ERIC}
}

@article{nguyen2025community,
  title={Community college articulation agreement websites: Students’ suggestions for new academic advising software features},
  author={Nguyen, David Van and Doroudi, Shayan and Epstein, Daniel A},
  journal={Community College Journal of Research and Practice},
  pages={1--17},
  year={2025},
  publisher={Taylor \& Francis}
}

@techreport{nscrc2024transfer,
  author = {{National Student Clearinghouse Research Center}},
  title = {Transfer and Progress},
  year = {2024},
  institution = {National Student Clearinghouse Research Center},
  url = {https://nscresearchcenter.org/transfer-and-progress/},
  note = {Accessed: 2025-05-05}
}

@article{chen2018unsupervised,
  title={Unsupervised multilingual word embeddings},
  author={Chen, Xilun and Cardie, Claire},
  journal={arXiv preprint arXiv:1808.08933},
  year={2018}
}

@article{artetxe2017unsupervised,
  title={Unsupervised neural machine translation},
  author={Artetxe, Mikel and Labaka, Gorka and Agirre, Eneko and Cho, Kyunghyun},
  journal={arXiv preprint arXiv:1710.11041},
  year={2017}
}

@article{lample2017unsupervised,
  title={Unsupervised machine translation using monolingual corpora only},
  author={Lample, Guillaume and Conneau, Alexis and Denoyer, Ludovic and Ranzato, Marc'Aurelio},
  journal={arXiv preprint arXiv:1711.00043},
  year={2017}
}

@article{reimers2019sentence,
  title={Sentence-bert: Sentence embeddings using siamese bert-networks},
  author={Reimers, Nils and Gurevych, Iryna},
  journal={arXiv preprint arXiv:1908.10084},
  year={2019}
}

@article{lloyd1982least,
  title={Least squares quantization in PCM},
  author={Lloyd, Stuart},
  journal={IEEE transactions on information theory},
  volume={28},
  number={2},
  pages={129--137},
  year={1982},
  publisher={IEEE}
}

@article{rumelhart1986learning,
  author = {Rumelhart, D. E. and Hinton, G. E. and Williams, R. J.},
  title = {Learning representations by back-propagating errors},
  journal = {Nature},
  volume = {323},
  pages = {533--536},
  year = {1986}
}

@article{stoet2020gender,
  author    = {Stoet, Gijsbert and Geary, David C.},
  title     = {Gender differences in the pathways to higher education},
  journal   = {Proceedings of the National Academy of Sciences},
  volume    = {117},
  number    = {25},
  pages     = {14073--14076},
  year      = {2020},
  publisher = {National Academy of Sciences},
  doi       = {10.1073/pnas.2002861117}
}

@article{jackson2020century,
  author    = {Jackson, Michelle and Holzman, Brian},
  title     = {A century of educational inequality in the United States},
  journal   = {Proceedings of the National Academy of Sciences},
  volume    = {117},
  number    = {32},
  pages     = {19108--19115},
  year      = {2020},
  publisher = {National Academy of Sciences},
  doi       = {10.1073/pnas.2003718117}
}

@book{russell2019human,
  title     = {Human Compatible: Artificial Intelligence and the Problem of Control},
  author    = {Russell, Stuart},
  year      = {2019},
  publisher = {Viking},
  address   = {New York, NY}
}

@article{ryu2013credit,
  title={Credit for prior learning},
  author={Ryu, Mikyung},
  journal={ACE Center for Policy Research and Strategy},
  year={2013}
}

@misc{californiaAB1111,
  title        = {California Assembly Bill 1111: Postsecondary Education: Common Course Numbering System},
  author       = {{California State Legislature}},
  year         = {2021},
  url          = {https://leginfo.legislature.ca.gov/faces/billNavClient.xhtml?bill_id=202120220AB1111},
  note         = {Chapter 568, Statutes of 2021. Accessed October 16, 2025}
}

@article{pataranutaporn2023influencing,
  title={Influencing human--AI interaction by priming beliefs about AI can increase perceived trustworthiness, empathy and effectiveness},
  author={Pataranutaporn, Pat and Liu, Ruby and Finn, Ed and Maes, Pattie},
  journal={Nature Machine Intelligence},
  volume={5},
  number={10},
  pages={1076--1086},
  year={2023},
  publisher={Nature Publishing Group UK London}
}

@report{crac2025ai,
  title        = {Statement on the Use of Artificial Intelligence (AI) to Advance Learning Evaluation and Recognition: Full Statement},
  author       = {{Council of Regional Accrediting Commissions (C-RAC)}},
  institution  = {Council of Regional Accrediting Commissions},
  year         = {2025},
  url          = {https://68d6c276-e98a-4e4c-a61e-d7c0a9aab604.usrfiles.com/ugd/68d6c2_437ddf23a8d649d5a78504b586911982.pdf}
}

@article{de2023psychological,
  title={Psychological factors underlying attitudes toward AI tools},
  author={De Freitas, Julian and Agarwal, Stuti and Schmitt, Bernd and Haslam, Nick},
  journal={Nature Human Behaviour},
  volume={7},
  number={11},
  pages={1845--1854},
  year={2023},
  publisher={Nature Publishing Group UK London}
}

@article{mikolov2013efficient,
  title={Efficient estimation of word representations in vector space},
  author={Mikolov, Tomas and Chen, Kai and Corrado, Greg and Dean, Jeffrey},
  journal={arXiv preprint arXiv:1301.3781},
  year={2013}
}

@article{bankins2024multilevel,
  title={A multilevel review of artificial intelligence in organizations: Implications for organizational behavior research and practice},
  author={Bankins, Sarah and Ocampo, Anna Carmella and Marrone, Mauricio and Restubog, Simon Lloyd D and Woo, Sang Eun},
  journal={Journal of organizational behavior},
  volume={45},
  number={2},
  pages={159--182},
  year={2024},
  publisher={Wiley Online Library}
}

@phdthesis{renaud2000examination,
  title={An examination of the barriers to articulation agreements between colleges and universities in Ontario},
  author={Renaud, Danielle},
  school={University of Toronto},
  year={2000}
}

\end{document}